# Examination of saturation coverage of short polymers using random sequential adsorption algorithm

Aref Abbasi Moud[1]*

[1] Department of Chemical and Petroleum Engineering, University of Calgary, 2500 University Dr NW, Calgary, AB, T2N 1N4, Canada

*Author to whom correspondence should be addressed; electronic mail: aabbasim@ucalgary.ca

**Abstract:** We filled a void with a regular or asymmetric pattern without overlap using a time-dependent packing method termed random sequential adsorption (RSA). In the infinite-time limit, the density of coverage frequently hits a limit. This study focused on the saturation packing of squares and their dimers, trimers, tetramers, pentamers, and hexamers, all of which were orientated in two randomly chosen orientations (vertical and horizontal). Our results concurred with those of previous extrapolation-based research[1]. We used the "separating axis theorem" to detect if freshly added polygons and previously put ones overlapped. When RSA insertion became disproportionately sluggish, we concluded that saturation had been attained. We also discovered that the system's capacity to fill the area decreased as squares were stretched into dimers and trimmers. The microstructure of the resultant saturation was also thoroughly investigated, including block function and the structural arrangement of dimers and trimers.

**Keywords:** RSA, short polymers, separating axis theorem, trimers, and dimers

## 1. Introduction

One of the most popular nonequilibrium packing models is the random sequential addition (RSA) packing method, which is a time-dependent process. Unpredictable sphere packings are created using a method similar to earlier established techniques; see References[2-3]. The RSA packing process in the three spatial dimensions[4] has been used to represent a range of scenarios, including protein adsorption[5], polymer oxidation, particles in cell membranes, and ion implantation in semiconductors[2].

Euclidean space (d-dimensional) particles with certain shapes are randomly and progressively introduced into the volume subject with the restriction that they do not overlap to carry out the simulation starting from a big, vacant zone. The freshly created particles are only maintained if they do not overlap any already existing particles; otherwise, they are deleted. After the simulation has begun, this procedure can be stopped at any time, making the density that has been acquired time dependent. Density reaches a "saturation" or "jamming" limit as time goes on[6].

In its simplest form, an RSA sphere packing may be obtained by randomly, irreversibly, and sequentially adding nonoverlapping objects into a huge volume that is initially empty of spheres. A further attempt is made until the sphere can be added without doing so if an attempt to add a sphere (or any other polygon) at time t overlaps with a sphere that is already present in the packing. An RSA configuration with a time-dependent packing fraction can be created by selecting any moment in finite time t as the process' endpoint. The maximum saturation packing fraction, which occurs at the infinite-time and thermodynamic limits, prevents this figure from becoming higher.

RSA implementations fall into two groups. The two fundamental categories into which the RSA models may be split are continuum and lattice models. Based on object kinds, they are further split into two groups: RSA of finite (nonzero) area objects and RSA of zero area objects. Things having finite area in this sense are those that enclose a certain amount of space, whereas those with zero area lack this geometric feature of enclosure. Therefore, upon adsorption on the substrate, items with finite area take up some space, whereas those with zero area take up no space. Lattice models are defined by the realisation of the jammed state[7], regardless of the item types.

In general, in case of RSA of finite area objects, the approach of instantaneous coverage $\theta(t)$ to the jammed state coverage $\theta_{max}$ is found to follow a power law $\theta_{max} - \theta(t) \sim t^{-p}$. Researchers have proposed certain laws about the value of the exponents p by researching the RSA of objects with a variety of geometries, including circular, elliptical, rectangular, and sphero-cylindrical. Feder [8] is credited with being the first to link the observed value of the exponent p in the RSA of circular objects in two dimensions to the object dimensionality and to propose the general rule that p = 1/D for d-dimensional circles on a two-dimensional continuum platform. Swendsen [9] subsequently demonstrated that, if the items are placed with random orientations, the same ought to apply for RSA of items of any arbitrary form.

Since its invention by Feder [8], Random Sequential Adsorption (RSA) has gained widespread acceptance as a technique for simulating adsorption characteristics, particularly for spherical molecules. However, employing RSA to replicate the adsorption of more complex particles like polymers or proteins raises concerns about how RSA's inherent features alter when non-spherical molecules are involved. For simple forms like spheroids, spherocylinders, rectangles, needles, and others, the subject has already been addressed[10-12]. Recent research, however, indicates that these geometries are insufficient for simulating the adsorption of common proteins, such as fibrinogen, for instance[13]. As a result, researchers' focus has recently turned to coarse-grained modelling of complex biomolecules and polymers[14-16].

In 1-D case, the saturation packing fraction can be obtained analytically as 0.747597920 [17] however for 2-D and 3-D the saturation packing fraction for discs and spheres has been estimated only through numerical simulations; the most precise ones are 0.5470735 ± 0.0000028 for 2-d and 0.3841307 ± 0.000002 for 3-D cases[2]. Other figures reported for 2-D simulations of discs are 0.5470735[2], 0.547067[18], 0.547070[19], 0.5470690[20], 0.54700[21], 0.54711[22], 0.5472[23], 0.547[8], 0.5479[24]. Similarly for 2-D aligned squares saturation coverage reported in the literature is 0.562009[25], 0.5623[24], 0.562[8], 0.5565[26], 0.5625[27], 0.5444[28], 0.5629[29], 0.562[30].

In this study, we used the RSA technique to determine the saturation packing limit for squares and its dimer and trimers. As the length of square increases, the results indicated that samples eventually generated structures with less and less packing. In this study, the "separating axis theorem" approach was used to detect whether there was a collision between two polygons; more information on the procedure is provided in the following sections. Polygons here refer to squares that encompasses its polymers as well as a constructing monomer. Our preliminary findings on RSA packing with respect to polygons, which we just published[31], are the basis for this study, which extends that work by extending the polygon (square) into its polymers.

## 2. Model and simulation procedure

When colloidal particles or molecules are being adsorbed, they frequently diffuse close to the surface. This process might lead to the formation of a film consisting of molecules that were randomly adsorbed because of adhesion. Here, we focus on adsorbate monolayers formed by irreversible adsorption. The most straightforward technique for quantitatively modelling these processes is molecular dynamics (MD). The advantages of MD include accurate forecasting and management of most environmental factors, such as temperature and the diffusion constant. The main issue is the performance deficiency. As a result, we decided to utilise a new method, continuum Random Sequential Adsorption (RSA), which has been successfully employed to investigate colloidal and other systems[32].

To simulate, a virtual particle was created (square, its polymers), and its location on an area was chosen at random based on a uniform probability distribution.

- The overlap with the previously adsorbed nearest neighbours of a virtual particle was tested (the topic of the next section). The result of this test tells us whether the surface-to-surface distance of a particle is greater than zero.

- If there was no overlap, the virtual particle was adsorbed and added to an existing covering layer.

- If there was overlap, the virtual particle was dropped and abandoned.

**2.1 Proposed algorithm**

Numerous methods may be used to determine if two polygons intersect or not. One method for determining if two polygons are overlapping uses mathematical equations and is known as the "separating axis theorem"[33].

The separating axis theorem states that if a line divides two convex polygons, they cannot intersect. The separation axis, which is a line, may be thought of as the normal to one of the edges of each polygon.

Using the separating axis theorem, the following procedures can be used to determine if two polygons cross:

- Determine each polygon A edge's edge normal, then project both polygons onto that value.

- Establish the minimum and maximum projections of each polygon onto the normal.

- If the maximum projection of polygon A is less than the minimum projection of polygon B, or the other way around, the polygons do not overlap.

- If the projections overlap, repeat the procedure for each edge in polygon B.

- If the projections of the polygons onto the separating axes do not overlap, the polygons are connected.

It is likely that non-convex polygons or polygons with holes will not be covered by this theorem, even though the separating axis theory may be used to determine if two convex polygons overlap. In some situations, it could be necessary to verify for intersection using alternative techniques.

**3. Results and discussion**

We construct saturated RSA configurations of polymers (dimer to hexamer) and compare saturation packing with other findings reported in the literature, particularly in ref[2], where authors employed a different approach and orientation was random, to show the precision and utility of our algorithm. We generate 1000 variations for each particle form using the system size that results in the fastest and densest packing.

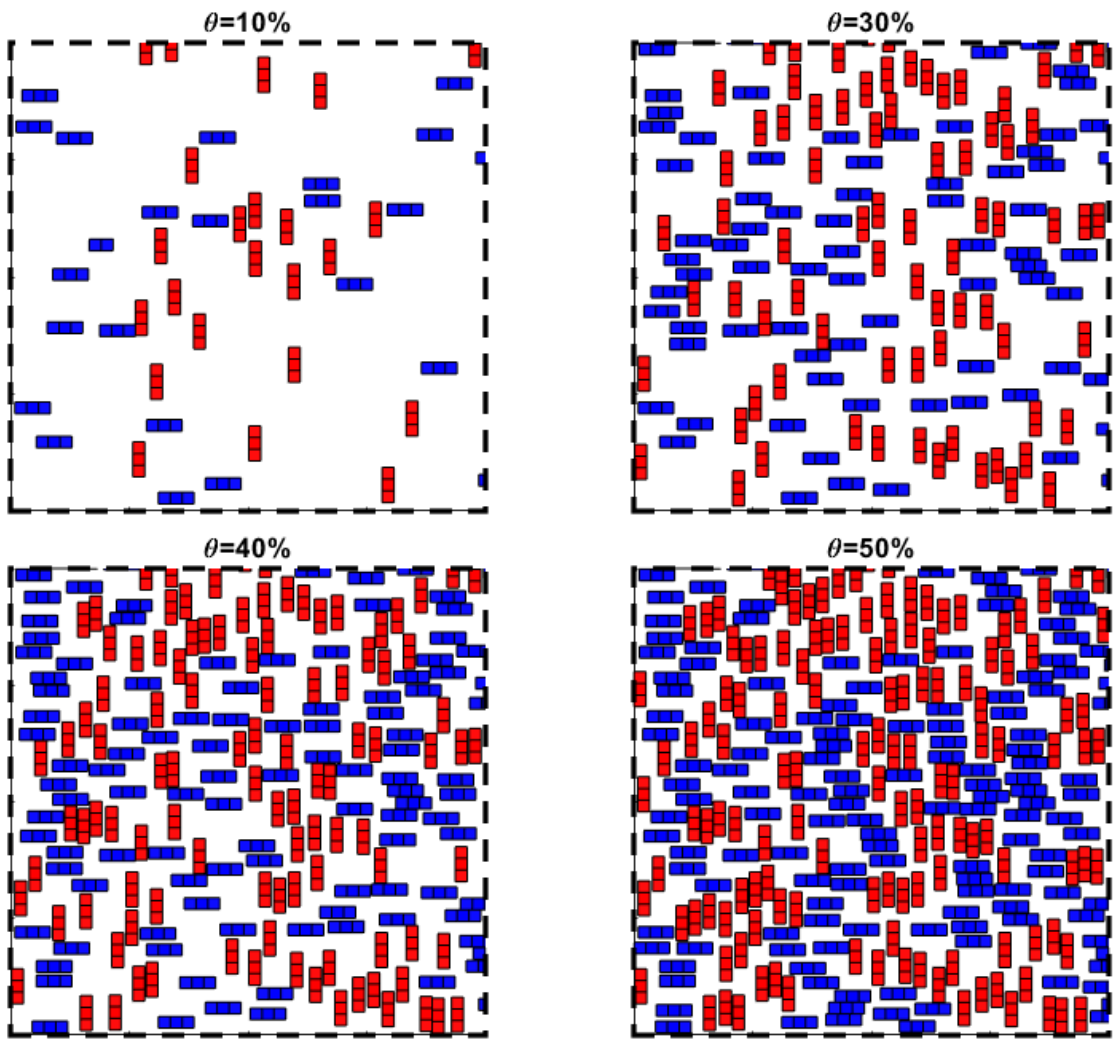

**Figure 1.** Typical monolayer samples (from trimer's sample) for three different coverages: θ = 0.1, θ = 0.3, θ = 0.4 and θ = 0.5 for trimers. The collector side length was equal to 50[-]. Fixed boundary conditions were used. Figure shows truncated images of the distribution for a better visibility (20 by 20).

Using the greatest area (50 by 50) with fixed bounds, most of the results presented later in the study were achieved. We made sure that the adoption of periodic boundary conditions had no discernible impact on the results that were made (Equivalent of periodic and fixed boundary condition). Figure 1 shows the outcomes of one of the simulations, in which trimers are placed in an area measuring 20 by 20 with a monomer having a side length of 0.5 units. Particles are gradually added to the surface as the simulation progresses.

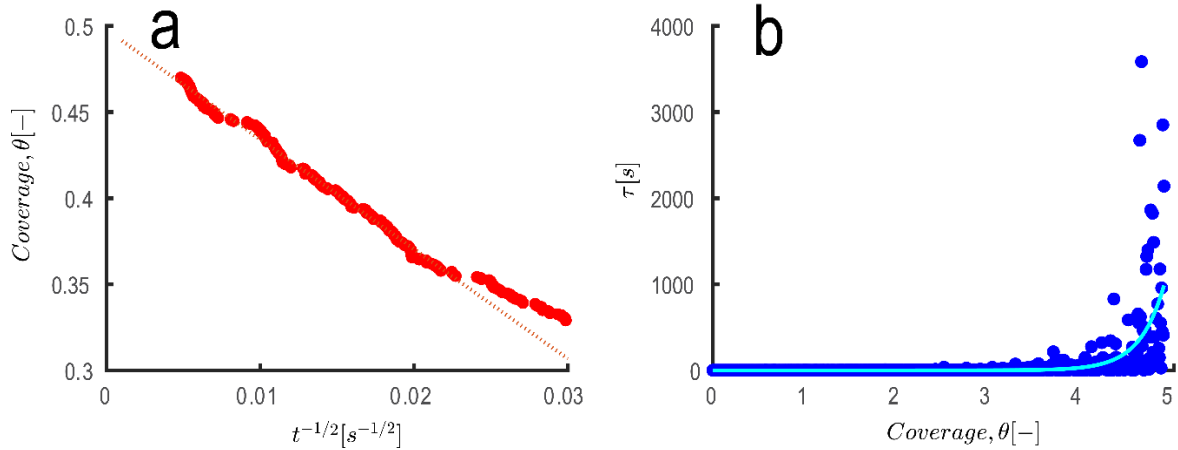

**Figure 2.** Hexamer units put irrevocably inside a 50 by 50 space are the focus of the RSA algorithm. **(a)** Asymptotic observation of coverage as a function of total simulation duration indicated by t. **(b)** The instantaneous time τ determined by based on coverage. The line represents an exponential fit.

To look at the development in more details, surface coverage was depicted as a function of simulation time to the power of $t^{-1/3}$ and results are shown in **Figure 2a**. Clearly at long simulation times of ~$10^4$ coverage very slowly reaches its asymptotic limit that is 0.4968±0.0011. Similarly using same schemes, we arrived at 0.5631±0.0002 surface coverage for squares. This surface coverage corresponds very well with the results reported in the literature for squares [8, 24-30].

The number of attempts needed to add a new particle to the grid (or collector both terms in congruency with literature has been used here interchangeably) can be known, and this information can be used to model the blocking of further adsorption through monitoring time. Clearly, as more of the surface is covered, adding new particles should be more difficult, which can be represented by a lower probability (See **Figure 2b**) that is it takes considerably more amount of time for a particle to be added. Adsorption kinetics in a real experiment typically depends on two variables: the effectiveness of the transport process (primarily diffusion or convection depending on the experimental setup) that moves the adsorbate from the bulk to the surface, and the likelihood of catching particles that are nearby [34-39]. Authors in other reports[40] have concentrated on the second aspect in this case, which is defined by the blocking function, also known as the available surface function (ASF). The simulation makes it simple to obtain it as a ratio of successful attempts to all RSA attempts. Equivalent to available surface function that is represented in shape of time simulation is presented in **Figure 2b**. **Figure 2b** shows simulation time as a function of coverage; statistically, it is evident that more trials are required to attain adsorption because the surface is already rather packed. The exponential fit is, $\tau = 0.0006 \exp(24.03\,\theta)$, thus describing increasing time required to place an additional point onto the grid. Discussion on ASF is subject of next section.

The main objectives of this work were to determine the maximum random adsorption ratio for squares and their polymers and compare it to the results for hard circles (spheres). That ratio ought to be provided for an infinite grid area and adsorption duration. Although one must deal with constrained simulation durations, one must also manage the accuracy issue brought on by the finite grid size. Because it is unclear if there would be any possibility of adsorption after the simulation time, particularly in the case of large grids, the determination of maximal coverage depends on the RSA kinetics model. As a result of prior research in the region [9, 41-42], there have been a number of works in the area, and asymptotically:

$$\theta_{max} - \theta(t) \sim t^{-\frac{1}{D}} \quad eq.1$$

Regarding the irreversible deposition of discs or squares that are not orientated (formerly known as p = -1/D). Despite controversy, D here specifies the grid's dimension[9]. When adsorbed particles are organised, the situation is altered[9, 42]. **Figure 2a** previously in this post showed an example of the results of fitting **Equation 1**. Asymptotic observations of coverage for squares seem to neatly match **Equation 1**. Although **Equation 1** hasn't been definitively proved, its validity has been vigorously defended[12] by analytical and numerical grounds. It should be noted that Equation 1 simplifies to the standard Feder's law[8] for isotropic objects since n equals the number of dimensions.

For instance, RSA of discs on a two-dimensional plane has D = 2, whereas RSA of rectangles, ellipses, and other rigid but noticeably anisotropic structures has D = 3 [37, 43]. It appears that parameter D generally correlates to the degrees of freedom of a number of packed objects, which has been validated for the random packing of hyperspheres in higher dimensions [2, 21], not only the integral ones[44-45]. The power law (**Equation 1**) is satisfied for the RSA of polymers examined here, however the exponent -1/D strongly relies on a polymer length. The parameter D is about equivalent to 3, which is the value recognised for anisotropic molecules, for a small number of vertexes such as pentagon and squares. However, as number of vertices increases parameter D converges to 2. This finding is consistent with those made for the RSA of spherical beads examined in Ref.[46]. However, unlike what was shown in the cases of spherical beads[46] or generalized dimers [40], there is no abrupt transition between these two limitations.

The results are averaged across 10 simulation runs with time t in the order of $5 \times 10^8$ for each run in order to determine parameter D for different polymers. These runs' data are not displayed here, and we will go into more depth about the outcomes in our upcoming paper.

### 3.1 RSA for polymers

In the last part, we laid the foundation for using the RSA approach to create oriented squares and trimmers. Results showing the behaviour of adsorption at asymptotic limits, the kinetics of the adsorption index (p), and the relationship between simulation time and coverage were given. Additionally, results and discs were compared. Utilizing the extrapolation method shown in **Figure 2a previously, Table 1 generates saturation densities for various** polymer lengths. **Figure 3** displays a sample of saturation densities for various forms.

To arrive at the values reported in **Table 1** following equation has been used:

$\theta(t) = \theta_{max} + b/t^p$    eq.2

When arriving at the values shown in **table 1,** we gave the data from longer simulation times more weight. The approach's possible downside is that each data point is given the same weighting factor, assuming all values are given the same weight. Because there are a lot more of these points in the higher part of the asymptotic area, it is sort of underweighted. Therefore, we investigated a novel strategy that introduces a bias favouring the longer durations. These changes are in line with the accounts in ref[12].

**Table 1.** Saturation density, index, for square-based polymers with a monomer to simulation box length ratio of 0.01 and their respective 95% confidence intervals.

| Shape (oriented) | $\theta_{max}$ [-] (95% confidence bounds) | p [-] (95% confidence bounds) | b [-] (95% confidence bounds) |
|---|---|---|---|
| Square | 0.5631(0.5629, 0.5633) | 0.5138 (0.5125, 0.5151) | -4.226 (-4.475, -4.561) |
| Dimer | 0.57 (0.5697, 0.5704) | 0.4599 (0.4595, 0.4603) | -4.805 (-4.819, -4.792) |
| Trimers | 0.5621 (0.5612, 0.5629) | 0.466 (0.4652, 0.4668) | -8.003 (-8.066, -7.941) |
| Tetramer | 0.5558 (0.5539, 0.5577) | 0.4998 (0.499, 0.5007) | -6.046 (-6.15, -5.942) |
| Pentamer | 0.5504 (0.5481, 0.5527) | 0.46 (0.4576, 0.4624) | -5.007 (-5.122, -4.892) |
| Hexamer | 0.4968 (0.4949, 0.4987) | 0.5 (0.499, 0.501) | -6.264 (-6.458, -6.069) |
| Discs | 0.547073[2] | - | - |

As outlined in introduction section, similarly for 2-D aligned squares saturation coverage reported in the literature is 0.562009[25], 0.5623[24], 0.562[8], 0.5565[26], 0.5625[27], 0.5444[28], 0.5629[29], 0.562[30]. Our values for square are very well within range of values reported elsewhere. However, as particles get longer and become Trimers, saturation has dropped since longer particles require more accessible area for deposition. For dimer and trimer, the greater aspect ratio of the dimer is projected to result in somewhat higher saturation for dimers. These findings are crucial because they suggest that it gets progressively harder for molecules to adhere to surfaces as they become longer; an example of superiority of simulation over experiment in giving researcher a tool to examine parameters hard to measure through experiments.

Results from this study can also be extrapolated to higher dimensions. For instance, the efficiency of a sequential adsorption process with hard materials decreases with increasing size. It is noteworthy to note that, as a general rule, the saturation coverage in D dimensions is very well estimated by that in one dimension raised to power D (for the RSA of spherical particles, $\theta_{max} \simeq 0.75$ for D = 1, $\theta_{max} \simeq 0.55$ for D = 2, $\theta_{max} \simeq 0.38$ for D = 3, etc)[47]. Therefore, results obtained here can be extended to 1-D and 3-D cases with good approximation, for instance for squares for cubes is predicted to lie around 0.38 and in 1-D case around 0.73.

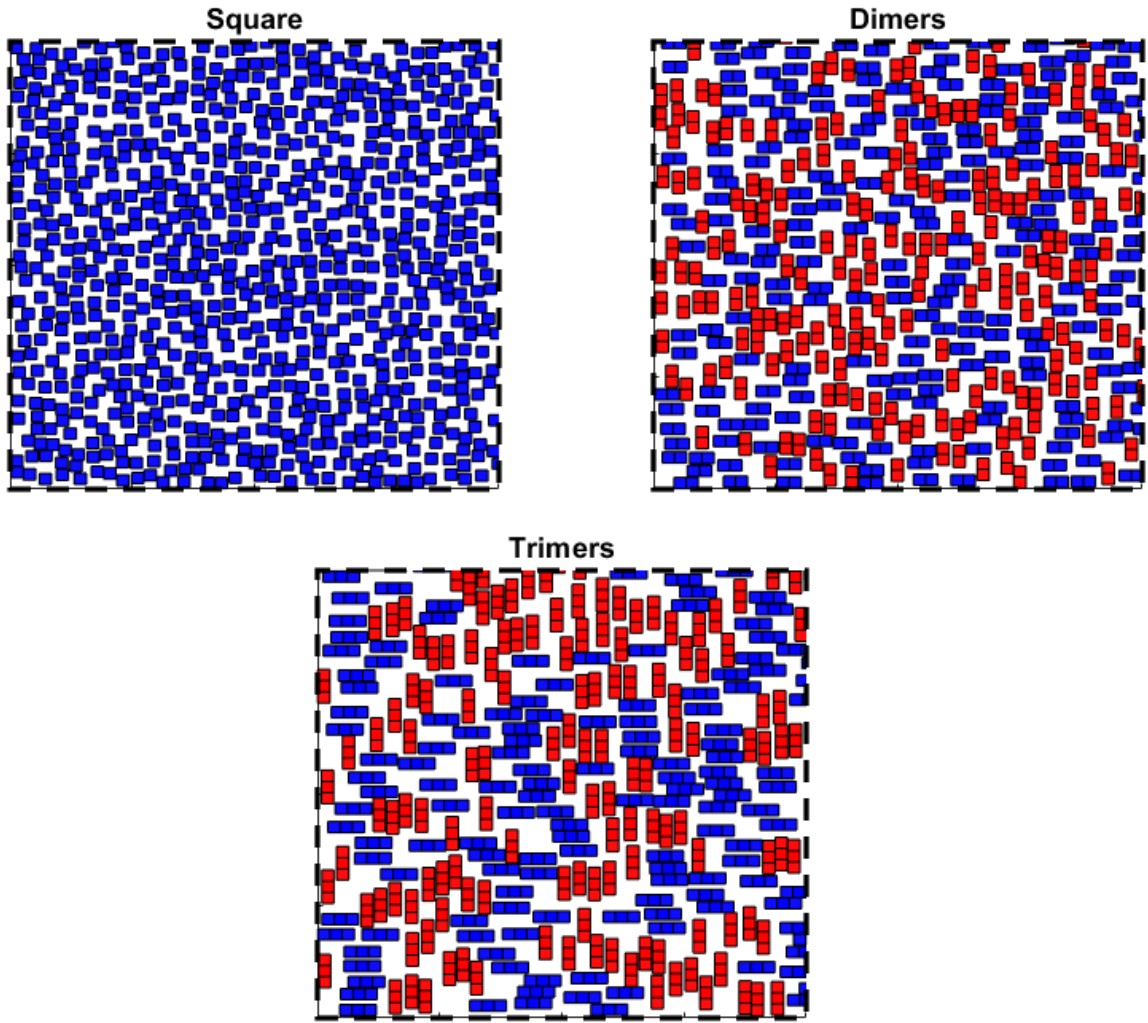

**Figure 3**. Square, dimers, and trimers near saturation points for samples with monomer's side length of 0.5 and distributed within area of 50 by 50. Fixed boundary condition has been applied. Figure shows truncated images of the distribution for a better visibility (20 by 20). For improved visibility, the horizontally oriented polymers have been coloured blue, while the vertically oriented ones have been painted red.

Clearly visually samples experience a bit higher coverage for dimers and less coverage for trimers.

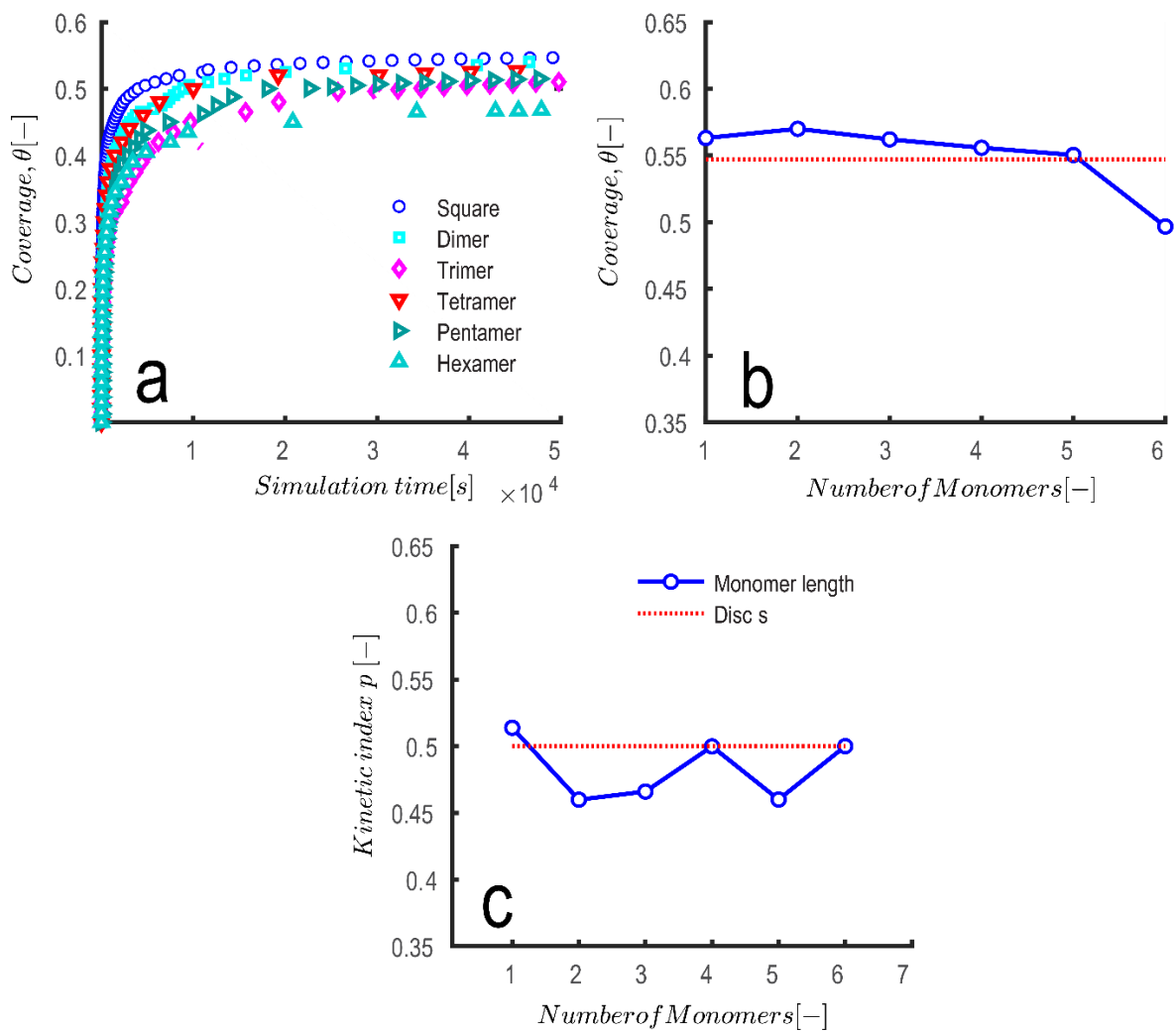

**Figure 4.** RSA saturation density for polymers, with a line created to guide the viewer's eyes. Each data point has an almost imperceptible error bar. For discs, RSA saturation density (red dotted line). All the polymers in **Table 1**'s saturation density changes as a function of simulation duration **(a)** The dependence of $\theta_{max}$ on length of the polymer **(c)** Kinetic index as a function of the monomer length.

**Figure 4** shows the RSA saturation density as a function of polymer length together with an eye-guiding line. The saturation density coverage first increases somewhat as the polymer's length rises, but as it continues to grow, it starts to decline. This outcome is in line with the outcomes for rectangles with various aspect ratios and ellipses that have been reported in the literature[5, 48]. Additionally, we discovered that saturation is somewhat lower for polymers with aspect ratios longer than 2, and we hypothesise that this is because, as was already noted, longer particles are more difficult to pack efficiently.

The random coverage ratio dropped exponentially with polymer size in earlier investigations refs[43, 49]. On a continuous surface, at least two competing variables can affect the maximum random coverage ratio. First, there is less chance of finding a large enough uncovered section to place on a collector, making it harder to separate larger particles than in the lattice case. The second point is that a polymer globule is more likely to form a cluster when necessary because it has a greater monomer packing ratio than a group of individual monomers. For continuous collectors as opposed to lattice ones, this second element is more important.

## 3.2 Block function

Knowing how many tries are necessary to add a new particle to the grid allows us to simulate how further adsorption is blocked over time. It is obvious that when more of the surface is covered, adding more particles should become more challenging, which may be represented by a decreased probability (as shown in **Figure 2b**, which depicts the same trend with time as the dependent variable)

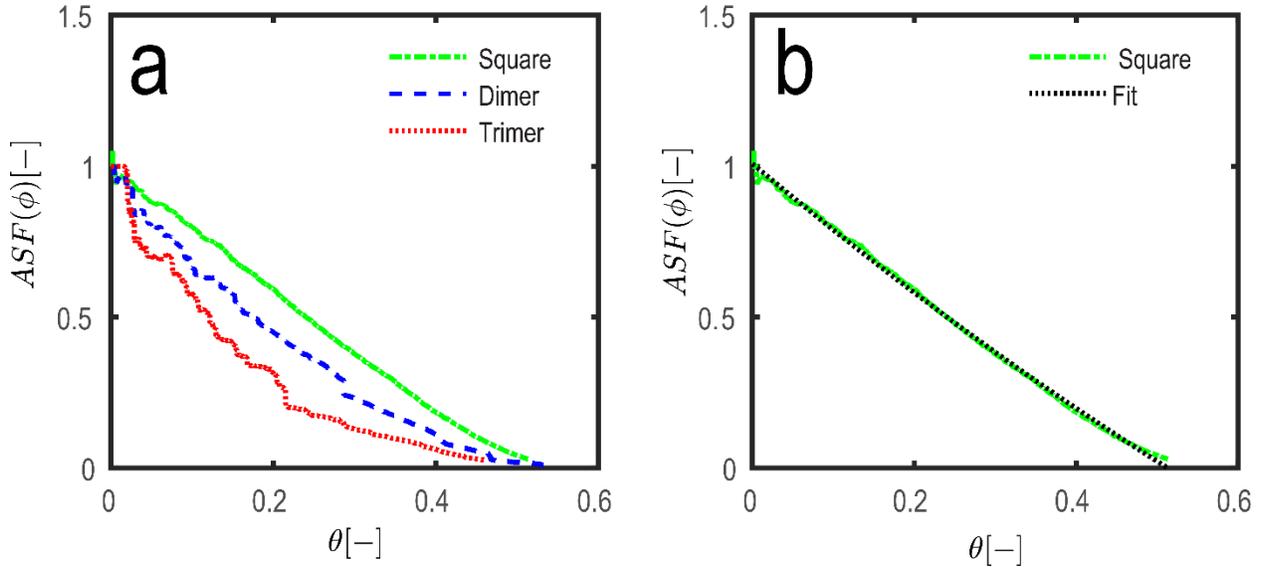

**Figure 5.** The ratio of successful attempts to blocking attempts versus coverage. The green line represents simulation data in **b** and the dotted line represent polynomial fit.; details of the fits are given in **table 2**.

**Equation 3** describe the function with a simple second order polynomial perfectly describing the decreasing trend in probability of a successful adsorption.

$$ASF(\varphi) = C1\theta^2 + C2\theta + C3 \; eq.3$$

In which C1, C2 and C3 are pre-factors. Details of the fit of equation 3 is given in the **Table 2. Figure 5** shows that the likelihood of adsorption for trimers drops more quickly than for dimers and squares, and the dimer with respect to the square exhibits the same pattern. This is because consecutive adsorption is much less likely to occur quickly in trimers and dimers than in squares due to their greater aspect ratio and unusual orientation.

**Table 2.** Represents polynomial fit to the data in **Figure 5** along with corresponding 95% confidence bounds.

| Shape (oriented) | C1 [-] (95% confidence bounds) | C2 [-] (95% confidence bounds) | C3 [-] (95% confidence bounds) |
|---|---|---|---|
| Square | 0.5439 (0.5296, 0.5582) | -2.244 (-2.252, -2.237) | 1.008 (1.007, 1.009) |
| Dimer | 2.508 (2.484, 2.532) | -3.18 (-3.194, -3.167) | 0.9796 (0.978, 0.9811) |
| Trimers | 5.483 (5.382, 5.585) | -4.463 (-4.511, -4.414) | 0.9623 (0.9574, 0.9671) |
| Tetramer | 5.359 (5.176, 5.543) | -4.338 (-4.437, -4.238) | 0.8921 (0.8809, 0.9034) |
| Pentamer | 6.352 (6.063, 6.641) | -4.828 (-4.94, -4.715) | 0.97 (0.9621, 0.9779) |

| Hexamer | 6.572 (6.359, 6.785) | -5.038 (-5.116, -4.961) | 0.9933 (0.9881, 0.9986) |

In the case of square, dimers, trimers simulations show that C1 = 0.5439-6.572 and C2 =-2.244-5.038, whereas those parameters for hard circles adsorption are analytically derived as C2 =-4 and C1 = 3.31[50] (we obtained coefficients of C1=2.426 (1.071, 3.782) and C2=-2.907 (-3.644, -2.169) with aid of our simulation) . Contrary to discs, the coefficient for squares suggests that they have an easier time adhering to the surface. Higher saturation coverage for squares is another manifestation of this phenomenon. Therefore, dimers values are very close to the values reported for hard discs.

According to **Figure 5**, as coverage levels increase, we get closer to the asymptotic phase, where dynamics are well understood, and finally the jamming limit. Thus, the RSA procedure has now been completely explained. Since it is challenging to conduct adsorption investigations near to the jamming limit[51], measuring the terms of the RSA process is the most effective technique to show that adsorption follows an RSA process. It's crucial to remember that words up to $\theta^2$ don't reveal anything about the properties of the adsorption process (i.e., the degree of irreversibility). This implies that any experiment that involves the adsorption of particles that resemble hard discs is susceptible to such an extension (to second order). Our strategy is also applicable to combinations and non-circular particles.

### 3.3 Ordering and orientation

We study the presence of any orientational order in a monolayer using the amorphous form of a polymer. Although most of these studies have used a collector surface lattice structure, as in ref. [52], such ordering has been well investigated. It could also have an impact on the RSA kinetics mentioned before. Based on the polymer structure, we offer the following function to measure orientational order in our continuous system:

$$S(\varphi) = \frac{1}{N}\sum_{i=1}^{N}(x_i \cos(\varphi) + y_i \sin(\varphi)) \quad eq.4$$

where (xi, yi) are positions along the i-th molecule in a layer for a unit vector. It is clear that $S(\varphi)$ is an average scalar product between both the orientation of molecules and the direction determined by an angle. As a result, for a perfectly aligned layer, $S(\varphi)$ will swing between 0 and 1, with highest values for angles parallel to molecules and minimum values for angles perpendicular to the alignment direction. $S(\varphi)$ will always be a constant and equal to 0.5 for pure random alignment.

For trimers as coverage increases across simulation time, ordering hovered 0.49, 0.53,0.50,0.51 as surface coverage increased from 10 to 30, 40 and 50%. Clearly ordering in trimer population is very close to random due to simulation being designed to give equal chance to parallel or vertical orientation of trimers as shown in **Figure 3**. Situation is very similar for dimers as well.

**Figure 4** illustrates how ordering in trimers may be further examined as a function of the radius of the particle neighbours. Figure 4 was made using a similar idea to the pair correlation function by treating the trimer centre as a circle with a radius of 0.25. $S(\varphi)$ fluctuates in small regions because dense clusters of horizontally or vertically oriented trimers are more likely to form, but as the sweeping radius grows, this fluctuation decreases to a value that is very similar to a randomly oriented arrangement. Dimers also face a similar set of circumstances. As a result, at r5, local order in each system vanishes. Small amplitude fluctuations continue after r=5, although their amplitudes and frequency are higher for trimers.

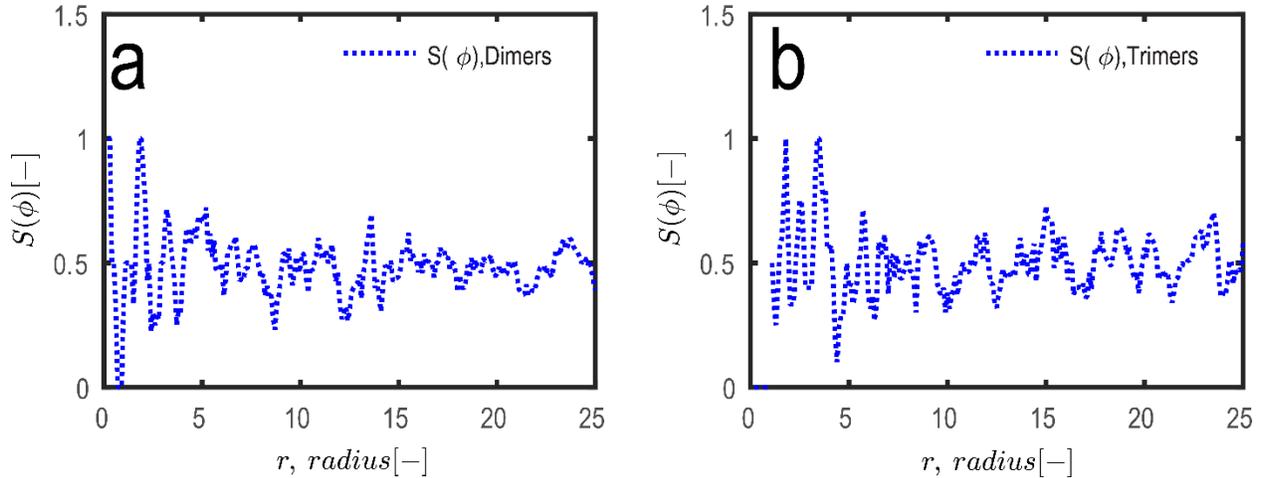

**Figure 4.** Local ordering as a function of radius for two RSA packings made with dimers and trimers near their saturation coverage. (a) dimers (b) trimers.

Similar trends are expected for tetramers due to similarity of behavior we have refrained from exploring them further here. As an example, liquid crystals are one type of orientationally organised structure that an elongated particle (such as polymers here with aspect ratio>2) can produce. When particle orientations are chosen at random using a uniform probability distribution for RSA on an infinite collector, the global orientational order is not expected to exist. However, since parallelly aligned particles take up less space, the formation of local ordered domains is feasible[53-54].

### 3.4 Radial distribution function

The radial distribution function (G(r)), also known as the pair correlation function, in a system of particles (such as atoms, molecules, colloids, etc.) explains how density changes in response to distance from a reference particle. G(r) is the radial distribution function [55] obtained from following equation:

$$G(r) = \frac{1}{\rho}\langle\sum_{i\neq 0}\delta(r-r_i)\rangle \ eq.5$$

The monolayer's first crucial characteristic is the particle autocorrelation. Squares are assumed to have a radius of 0.25 and to be treated equally regardless of whether they are made of the same polymer or a different one in order to compute G(r). **Figure 5** displays the average structures seen by various RSA packings. We consistently saw a peak at a distance of r=0.5 (right on the edge of the particles). In other words, the function reaches its maximum for the closest neighbour, r = 0.5, and then begins to degrade because of the volume that is lacking. The similar trend is seen in the trimer and hexamer, but there are more peaks. Hexamer contains additional peaks, for instance, at r=1, 1.5, and 2, while the trimer exhibits an additional peak at 1.

As the radius gets bigger, these peaks get weaker. Due to the coverage's randomness, these oscillations superexponentially vanish[21], and after normalisation, the function stabilises at a value of 1. In addition, when the number of monomers inside the polymer rises, the first peak corresponding to the nearest neighbour grows progressively sharper. According to this behaviour, particles that may be seen as a chain of squares pack more well even if the saturation coverage is smaller for monomers (squares) and short oligomers (dimers).

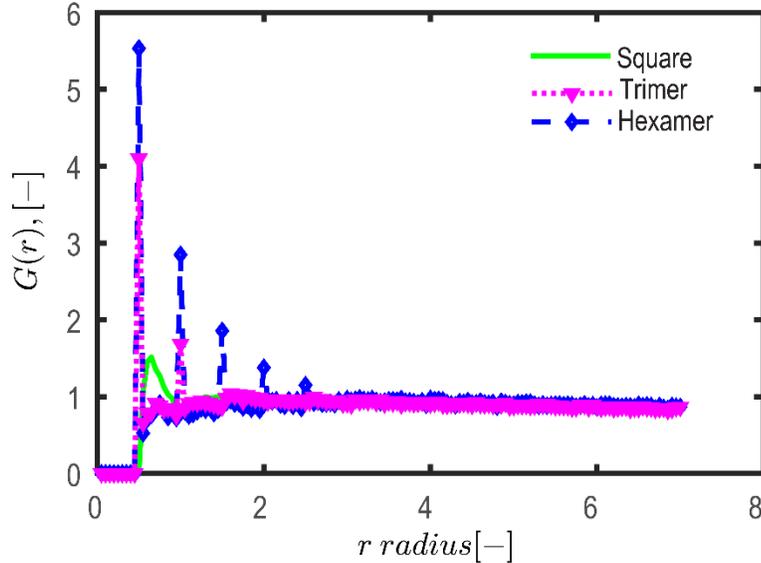

**Figure 5.** Functions of autocorrelation. Behavior autocorrelation function is depicted as a function of radius for square, dimer, trimer, tetramer, pentamer and hexamer.

The average structure seen by a generic particle of the system described by G(r) displayed in **Figure 5**, shows a full agreement with the predicted theoretical regimes found in literature [56-57]. In all cases, we observe a pronounced peak at a distance r~0.5, with the sphere diameter that corresponds to the distance of the nearest neighbors in contact. For r larger than the diameter, the probability to find neighbors decreases. In fluid-like systems, theoretically for $\varphi \lesssim 0.55$, [56-58] the G(r) is known to oscillate with decreasing amplitude.

**Conclusions**

For a range of stiff polymers produced using squared monomers, we show the maximum random coverage or saturation coverage in this paper and contrast our findings with those reported in the literature. In order to do this, we enhanced an algorithm that was described in Ref [32]. We prove the validity of our method by calculating the RSA saturation densities of polymers (dimer, trimer, and tetramer) and showing their consistency with prior findings in the literature.

The RSA model shown here may be extended to squares that may change into rectangles with larger aspect ratios to incorporate anisotropic particles in future research. Moreover, like ref[43] it can also include branched or more flexible polymers. Biological molecules are usually non-spherical, as seen by the previous example, and when their surface area in contact with the substrate is greatest, they firmly cling. According to experimental results, Schaaf et al. [59] discovered that the maximum substrate coverage they were able to achieve during the adsorption of fibrinogen—a non-spherical protein with an aspect ratio of roughly 7.5— was only about 40%, which was lower than the absorption coverage predicted by the RSA of hard discs— which is around 55%—and seen in experiments involving reasonably spherical globular proteins[8] (A similar impact was noted for albumin adsorption [60] ).

Here are some pertinent queries:

How does increasing the aspect ratio affect the saturation coverage of the substrate?

How does the particle shape impact the kinetics over both short and long time periods?

What are the similarities and differences between equilibrium configurations produced by RSA and configurations with equivalent surface coverage?

These questions will get their solutions in upcoming publications. This study's findings are pertinent since they considered a variety of particle morphologies, including those of asphaltene, graphene, cellulose nanocrystals, and kaolinite, among others [61-64].

The findings are important because they might help to understand how polymers behave when they are close to surfaces. For instance, numerous biological processes depend on proteins adhering to different surfaces. Understanding and having control over how protein molecules attach to surfaces and interact with them is essential when creating biomaterials. For instance, among other things, the production of biocompatible materials requires decreasing the adsorption of blood proteins to the material's surface. It is generally known that platelet adhesion followed by blood protein adsorption can result in surface-induced thrombosis. When protein adsorption is prevented or diminished, there is very little platelet adhesion to the surface. Eliminating lysozyme buildup from the surface of contact lenses is another illustration. In other circumstances, we would like to promote the adsorption of some proteins while inhibiting the adsorption of others.

**Conflict of interest statement:** Author declares no conflict of interest

**Data availability statement:** The datasets generated during and/or analysed during the current study are not publicly available but are available from the corresponding author on reasonable request.